# Key role of lattice symmetry in the metal-insulator transition of NdNiO$_3$ films

Jack Y. Zhang, Honggyu Kim, Evgeny Mikheev, Adam J. Hauser, and Susanne Stemmer*

*Materials Department, University of California, Santa Barbara, CA 93106-5050, USA*

\* Corresponding author. Email: stemmer@mrl.ucsb.edu

**Abstract**

Bulk NdNiO$_3$ exhibits a metal-to-insulator transition (MIT) as the temperature is lowered that is also seen in tensile strained films. In contrast, films that are under a large compressive strain typically remain metallic at all temperatures. To clarify the microscopic origins of this behavior, we use position averaged convergent beam electron diffraction in scanning transmission electron microscopy to characterize strained NdNiO$_3$ films both above and below the MIT temperature. We show that a symmetry lowering structural change takes place in case of the tensile strained film, which undergoes an MIT, but is absent in the compressively strained film. Using space group symmetry arguments, we show that these results support the bond length disproportionation model of the MIT in the rare-earth nickelates. Furthermore, the results provide insights into the non-Fermi liquid phase that is observed in films for which the MIT is absent.

Understanding and controlling metal-to-insulator transitions (MITs) of correlated transition metal oxides has been a longstanding goal in materials physics [1,2]. Identifying the microscopic mechanisms of these MITs is difficult, as various electronic, structural, and spin degrees of freedom can compete or cooperate. As a result, the subject continues to be a topic of significant controversy even for seemingly simple materials. A prime example is the rare-earth nickelates ($R$NiO$_3$, where $R$ is a trivalent rare earth ion, but not La), which are prototypical strongly correlated materials that undergo an MIT as the temperature is lowered. In addition to the temperature-driven MIT, understanding the interplay between atomic structure and electronic degrees of freedom is key to other fascinating aspects of these materials, such as non-Fermi liquid phases [3-5] and the tunability of their Fermi surface to mimic that of the cuprate superconductors [6,7].

$R$NiO$_3$s adopt the distorted GdFeO$_3$-perovskite-derived structure (space group *Pbnm*). With decreasing rare-earth ion size, the structural distortion increases, shifting the MIT to higher temperatures [8]. Hydrostatic pressure [5,9,10] or epitaxial film strain [4,11-16] can also be applied to modify the MIT temperature ($T_{MIT}$). The direct correlation between $T_{MIT}$ and ionic radius (or strain) suggests that the Ni-O-Ni bond angles play a role, as they determine the electronic bandwidth and magnetic exchange interactions. Early studies thus identified the MIT in the $R$NiO$_3$s as resulting from a bandwidth controlled charge transfer gap [8,17] with concomitant orbital ordering to describe the antiferromagnetic ground state [18]. More recently, however, subtle symmetry changes from orthorhombic *Pbnm* to monoclinic *P*2$_1$/*n* have been detected at $T_{MIT}$, pointing to a charge ordered ground state, and a lifting of the degeneracy of the singly occupied $e_g$ band [19-21]. The orthorhombic-to-monoclinic transition permits two inequivalent Ni sites. It should be noted, however, that the ordered ground state may be more correctly described as a bond length ordered state, as it is characterized by alternating NiO$_6$ octahedra with different Ni-O bond lengths [21,22], and the nominal charges on the Ni sites may not be very different. For simplicity, and since this study focuses on the underlying lattice symmetry, we use the various terminologies (charge/bond order, charge/bond length disproportionation) interchangeably here, following most of the literature. While these new findings offer structural parameters for theoretical calculations, important questions remain as to the relative importance of the electronic, magnetic, and structural parameters driving the MIT.

To this end, thin film heterostructures have been proposed as a means for separating the lattice from the electronic and magnetic effects. For example, strain in thin films can affect the octahedral tilt patterns in LaNiO$_3$ films [23,24] and, concomitantly, their electronic structure, including band degeneracy [25]. Meyers et al. did not observe a symmetry change in x-ray absorption spectroscopy and resonant x-ray scattering of ultrathin (15 unit cells) NdNiO$_3$ films as they undergo an MIT, concluding that the magnetic ordering drives the MIT and that it can be decoupled from the structural distortion [26]. Upton et al. also dismiss charge order or symmetry changes, and propose Ni 3$d$ hybridization with O 2$p$, as well as Ni charge redistribution to Nd 5$d$ states as underlying mechanisms [27]. These results are in contrast to experimental [22,28,29] and theoretical works [30-32] that associate the MIT with charge/bond length order on the Ni sites. The resolution of this debate hinges on the ability to detect very subtle symmetry changes in strained $R$NiO$_3$ films as they undergo the MIT. The monoclinic distortion in bulk NdNiO$_3$ is very small [19] and would be even more challenging to detect in ultrathin films [33]. Furthermore, key insights could also be obtained from the opposite case, namely identifying the microscopic reason(s) when the MIT is *absent*, such as for compressively strained films.

In this work, we use position averaged convergent beam electron diffraction (PACBED) in scanning transmission electron microscopy (STEM) to analyze the structure of strained NdNiO$_3$ films grown on NdGaO$_3$ (tensile) and YAlO$_3$ (compressive). PACBED is sensitive to extremely small structural distortions [34], even if they only occur on the oxygen sublattice [24,35], and has a high spatial resolution. The overlapping diffraction discs create unique patterns, characteristic of the point group symmetry (or more accurately, Laue symmetry) of the structure. We obtain PACBED patterns above and below $T_{MIT}$. We show that NdNiO$_3$ films on NdGaO$_3$, which exhibit an MIT, undergo a structural transition while NdNiO$_3$ films on YAlO$_3$, which are metallic at all temperatures, do not. The results, combined with symmetry arguments, provide a complete and remarkably simple understanding of the MIT and its suppression.

15 unit cell (~ 6 nm) NdNiO$_3$ films were grown by RF magnetron sputtering on NdGaO$_3$ and YAlO$_3$ substrates, in an Ar/O$_2$ gas mixture, with a 9 mTorr growth pressure, as described in detail elsewhere [15]. Neither film is relaxed. The in-plane longitudinal resistivity was measured as a function of temperature in a Quantum Design Physical Properties Measurement System (PPMS). TEM cross-sections along [001]$_O$ and [1$\bar{1}$0]$_O$ (the subscript indicates the orthorhombic orientation) were prepared using a focused ion beam with final milling energies of 5 kV Ga ions. High-angle, annular dark-field (HAADF)-STEM imaging and (LA-)PACBED experiments were conducted on a 300 kV FEI Titan S/TEM ($C_s$ = 1.2 mm). A convergence semi-angle of 9.6 mrad was used for high resolution STEM imaging, while 9.6 and a reduced angle of 3.4 mrad was used for PACBED. LA-PACBED patterns are obtained from roughly a 12×12 unit cell area. An FEI double-tilt holder was used for room temperature PACBED and high resolution imaging, while a Gatan 636 double-tilt LN$_2$ holder was used for low temperature experiments. All cold-stage experiments were carried out a temperature of 105 K, which remained stable throughout the data acquisition. PACBED simulations were carried out using the Kirkland multislice code [36] at 0 K.

Resistivity curves as a function of temperature for the films studied here are shown in Fig. 1, along with corresponding STEM images of the film grown on NdGaO$_3$ at room and cryo temperatures, respectively. Both films were 15 pseudocubic unit cells thick and coherently strained to the substrate (0.85% tensile and -3.6% compressive for NdGaO$_3$ and YAlO$_3$, respectively), as verified by high-resolution x-ray diffraction [15]. While the film grown on YAlO$_3$ remains metallic down to the lowest temperature, an MIT at ~150 K, with a hysteresis of ~25 K, is observed in the film on NdGaO$_3$. $T_{MIT}$ is comparable to films reported in the literature on this substrate and with this film thickness [26], but lower than that of bulk (~200 K [8]), as is typical for $R$NiO$_3$ films. The cold stage temperature of 105 K is well within the insulating regime. The complete suppression of the MIT in films under compressive strain for this film thickness is consistent with reports in the literature [4,14,37].

Figure 2a illustrates the orientation relationship of an orthorhombic film grown on a (110)$_O$ orthorhombic substrate surface, which was confirmed in transmission electron microscopy (TEM). While a (001)$_O$//(110)$_O$ orientation would have a similar lattice mismatch, interfacial oxygen octahedral connectivity issues make this orientation unlikely. The film is epitaxially constrained along [1$\bar{1}$0]$_O$ and [001]$_O$. The orthorhombic $a$ and $b$ lattice parameters change to accommodate the strain, resulting in a characteristic angle $\gamma$ that deviates from the 90° angle in the bulk. A 2×2×2 pseudocubic supercell containing the original orthorhombic unit cell is outlined in blue. TEM cross-sections accessed both the [001]$_O$ and [1$\bar{1}$0]$_O$ projections, each with distinct features in the PACBED pattern, which are similar to those previously observed in orthorhombic films grown on cubic substrates [38]. A color table has been applied to all patterns in the present study, to better highlight the intensity changes. PACBED patterns along $[1\bar{1}0]_O$ contain a square-like feature in the central disc, while those along $[001]_O$ show a diagonal intensity stripe, as shown in Fig. 2b. These features are reproduced in PACBED simulations, and experimentally observed in other perovskite structures with the *Pbnm* space group (e.g. $NdGaO_3$, $YAlO_3$, $GdTiO_3$). They allow for identification of the zone axis by simple visual inspection. The present paper focuses on the $[001]_O$ projection (orange in Fig. 2a), since the PACBED patterns along $[1\bar{1}0]_O$ are less sensitive to the symmetry changes that are relevant here. While PACBED has been shown to be sensitive to small structural distortions [24,34,35], we also employ low-angle PACBED (LA-PACBED) to better distinguish the diffracting discs, and gain additional insight into the film structure. The reduction in semi-convergence angle used for LA-PACBED results in a decrease in the resolution of the corresponding high-angle annular dark-field (HAADF) STEM image that is acquired in parallel. However, the 3.4 mrad convergence angle remained sufficient to differentiate the film and substrate during LA-PACBED acquisition.

Figure 3a shows experimental, room temperature LA-PACBED patterns from the substrate and film regions of $NdNiO_3/NdGaO_3$, with a simulated orthorphombic $NdNiO_3$ bulk pattern (simulated at 0 K for speed) for comparison. A Sobel edge filter, which highlights sharp changes in intensity, was applied to each pattern and displayed in Fig. 3b, with relevant diffraction discs indexed in the orthorhombic notation. Insets of the 200 and 020 discs for the experimental patterns are also highlighted in Fig. 3b. In the room temperature measurements, we note a characteristic bright band in the LA-PACBED patterns that runs diagonally (bottom left to upper right) in the central disc of the simulated bulk $NdNiO_3$ and substrate $NdGaO_3$. A similar feature appears also in PACBED (Fig. 2b). Furthermore, we see clear differences in diffraction features between 200 and 020 discs: a bright region of intensity near the central beam, as well as 310 reflections, which can be clearly seen in the 200 discs, but are barely observed in the 020 discs. These asymmetric features are a close match with the bulk NdNiO$_3$ simulation and are caused by the octahedral rotations and A-site cation displacements of the orthorhombic *Pbnm* structure.

In contrast, the LA-PACBED pattern of NdNiO$_3$ on NdGaO$_3$ does not show the asymmetry in the intensity between the 200 and 020 discs, indicating that the strained film has different symmetry than the orthorhombic substrate. In addition, the 200 and 020 disc spacings, corresponding to the *a* and *b* orthorhombic lattice parameters, are clearly different for the NdGaO$_3$ substrate, as expected, but *a=b* for the NdNiO$_3$ film (although we note that the bulk *a* and *b* lattice parameters of NdNiO$_3$ are almost identical as well). This result is expected for a tensile strained orthorhombic film [39]: as the film is constrained along the $[1\bar{1}0]_O$ and $[001]_O$ directions, the strain is accommodated by an increase in the angle γ and a reduction in octahedral tilts along the growth axis (Fig. 2a). We note that this structural change in the film is a result of the strain state, irrespective of the symmetry of the substrate (i.e. cubic or orthorhombic). A measurement of γ from Fig. 2b yields γ = 90.2° and 93.2° for the NdGaO$_3$ substrate and NdNiO$_3$ film, respectively. These results indicate that the octahedral tilts about the $[110]_O$ direction are either mostly or completely suppressed.

Figure 3c illustrates the most likely Glazer tilt configuration [40] of the tensile strained NdNiO$_3$ film, which is the same as previously found by Vailionis [39]. Here, we use the convention of denoting the axis of zero tilt as the *a*-axis. The phase of the rotations remains the same as in the bulk, but the magnitudes change, with negligible tilt along $[110]_O$. This space group is more accurately described as orthorhombic *Cmcm*, and the in-phase and out-of-phase tilts are likely different. In particular, the in-phase tilt is probably small. LA-PACBED simulations (not shown) indicate that the similarities in intensity between 200 discs from Fig. 2a, as well as the absence of 120 reflections, are likely indicators of significantly reduced octahedral tilts along the projected direction ($[001]_O$). While the present study does not represent a rigorous determination of the space group symmetry based on electron diffraction [41], due to experimental challenges (thinness of our samples and stability issues), it still presents compelling evidence of the structural symmetry and octahedral rotations that are consistent with previous results from literature [39] and geometric arguments of allowed tilts within the respective space groups [42].

Figure 4 shows LA-PACBED patterns of the same sample at 105 K. While the substrate pattern shows similar features as the room temperature pattern (expected since NdGaO$_3$ does not undergo a structural transition [43]), we see a noticeable difference in the NdNiO$_3$ film. Most markedly, there is a strong intensity asymmetry between 110 and 1$\bar{1}$0 reflections, which can be seen as bright overlaps in the central disc. This difference indicates a reduction in symmetry, most likely a monoclinic transition to $P2_1/n$, similar to what occurs in the bulk. In particular, similar peak splittings between 40$\bar{4}$/404 reflections were observed by synchrotron powder diffraction in polycrystalline NdNiO$_3$, and taken as a sign of the monoclinic transition [19]. At low temperatures, $\gamma$ decreases to 91.1°, indicating an increase in octahedral rotations about the growth direction ($[110]_O$), consistent with $P2_1/n$.

Figure 5a shows LA-PACBED patterns of compressively strained NdNiO$_3$ grown on YAlO$_3$. These films are expected to be monoclinic ($P2_1/m$) [39] in this strain state. Patterns from the substrate (not shown) were similar to the NdGaO$_3$ substrate patterns at both temperatures. From Fig. 5a, we do not observe any noticeable differences between the room temperature and 105 K patterns, although the 220 and $2\bar{2}0$ reflections may have slight differences in intensities.  The NdNiO$_3$ film grown on YAlO$_3$ contained structurally disordered regions in HAADF STEM, likely as a result of the very large compressive strain that makes it more susceptible to TEM sample preparation and beam damage, which might explain overall weaker diffraction intensities.  A schematic of the most likely octahedral tilt rotations of the compressively strained NdNiO$_3$ is shown in Fig. 5b. The angles, $\gamma$ and $\alpha$, from the figure are consistent with angle measurements from the LA-PACBED pattern ($\gamma = 88.1°$ and $\alpha = 89.4°$). The tilt configurations for both compressive and tensile strained films are determined from geometric considerations based on the lattice parameters of the film and the substrate epitaxial constraints [39,42,44].  A detailed explanation for determining the tilt systems of similar films can be found in ref. [39].

At first glance, it may seem curious that the compressively strained film, which we believe to be monoclinic and have a similar tilt pattern as bulk NdNiO$_3$ (Figs. 3c and 5b), does not undergo an MIT.  To explain this observation, we first highlight the fact that the expected space group of the strained compressive film ($P2_1/m$) is *not* the same as the low temperature bulk NdNiO$_3$ space group ($P2_1/n$).  In general, the presence of order, i.e., charge or bond length disproportionation, will always result in a loss of symmetry, usually involving a loss of translational symmetry since neighboring octahedral sites are no longer equivalent.  As discussed by Woodward [42], order in the *Pbnm* space group results in a symmetry reduction to $P2_1/n$, and this is indeed observed in bulk nickelates.  Likewise, introducing 1:1 order into the $P2_1/m$ space group would reduce the symmetry of the unit cell to $P\bar{1}$ (triclinic).  However, any epitaxially strained film on a cubic or (110) orthorhombic substrate (90° in-plane angle) is bound to contain higher symmetry elements than the triclinic system.  Therefore, by symmetry arguments, compressively strained films are unable to reduce to a charge or bond ordered state upon cooling. These simple considerations, which are consistent with the charge/bond length order driven MIT, explain why these films remain metallic and do not undergo a MIT.

Meanwhile, Figs. 3 and 4 show clear evidence that the MIT in tensile ultrathin NdNiO$_3$ films is accompanied by a symmetry-lowering structural distortion. They likely belong to the $P2_1/n$ space group, which is consistent with our observation that octahedral tilts along the growth direction are re-introduced. In addition, while the tensile-strained film, grown on a [110]$_O$ substrate, is orthorhombic, we note that tensile strained films grown on cubic substrates, such as (LaAlO$_3$)$_{0.3}$(Sr$_2$AlTaO$_6$)$_{0.7}$ (LSAT) or SrTiO$_3$, would contain a tetrad axis along the growth direction, and therefore possess tetragonal symmetry. While seemingly trivial, the most likely Glazer tilt pattern with a tetragonal space group would be $a^0b^+b^+$, meaning the out-of-phase tilt in Fig. 3c would change to in-phase. Although such a change might seem to have a large effect on the transport properties, transport measurements [4] suggest otherwise: films grown on LSAT and SrTiO$_3$ show MITs, which shift to slightly higher temperatures with increasing tensile strain. These observations all support the view that the symmetry lowering structural distortion is a key requirement for the MIT; the exact high temperature symmetry starting structure is not as important *as long as it permits a transition to a lower-symmetry ordered state*. This is possible in case of the tensile strained films but not for the compressive strained films.

In summary, we have shown that LA-PACBED allows for the detection of subtle symmetry changes in ultrathin films due to epitaxial film strain and the MIT, which may be missed in other diffraction methods. Epitaxial film strain affects the high-temperature (above the MIT) octahedral rotations and space group symmetry. Tensile strained NdNiO$_3$ is best described with a larger centered unit cell (*Cmcm*) with Glazer tilt pattern $a^0b^+c^-$, while compressively strained films ($P2_1/m$) retain the [110]$_O$ out-of-phase rotations and have tilt patterns $a^+b^-c^-$, with octahedral tilts similar to those of bulk NdNiO$_3$ ($a^+b^-b^-$). Thus even at room temperature, strained NdNiO$_3$ films are structurally dissimilar from their bulk counterparts, and the modified structure, rather than the bulk P*bnm* space group, should be used as the starting structure in future theoretical work describing the MIT of coherently strained films and associated phenomena, such as orbital polarization and Fermi surface tuning. Furthermore, the results provide a remarkably simple understanding of the modified MIT in thin films: in the case of tensile strained film, transition to a symmetry consistent with charge order is allowed by the high-temperature space group, thus permitting the MIT, while the opposite is true for compressively strained films which therefore have to remain metallic.

More broadly, the results present compelling evidence for charge/bond length order being inextricably linked to the insulating state of the *R*NiO$_3$s, the transition to which we have shown to be highly reliant on the high temperature "parent" space group symmetry. The ability to tune the MIT by epitaxial film strain, in conjunction with sensitive measurements of the film symmetry, provides strong evidence that the charge or bond length ordered state is firmly linked to the MIT of the nickelates, which does not occur without it.

Finally, we note that the results also provide insights into the nature of the non-Fermi liquid phase that is observed in *R*NiO$_3$ films and bulk materials for which the MIT is suppressed [4,5]. In particular, the results show that the non-Fermi liquid phase in the thin films coincides with a phase whose lattice symmetry is incompatible with reaching the long-range ordered state. It would be interesting to determine if a similar mechanism, namely a suppressed or frustrated symmetry-lowering ordered ground state, may explain the appearance of non-Fermi liquid phases in other correlated systems.

**Acknowledgements**

We thank Nelson Moreno for help with the film growth. The microscopy work was supported by the U.S. Department of Energy (grant number DEFG02-02ER45994). The film growth and transport measurements were supported by FAME, one of six centers of STARnet, a Semiconductor Research Corporation program sponsored by MARCO and DARPA. Facilities used in this work were supported by the UCSB Materials Research Laboratory, an NSF-funded MRSEC (DMR-1121053).

**Figure Captions**

**Figure 1:** Resistivity as a function of temperature for NdNiO$_3$ films grown on NdGaO$_3$ (solid line) and YAlO$_3$ (dashed line) substrates. A MIT occurs on the NdGaO$_3$ grown film at ~130 K. Corresponding STEM images for the film grown on NdGaO$_3$ at room and cryo temperatures are shown on the right. Arrows mark the approximate interface between the film and substrate. Images were acquired using fast acquisition and cross-correlated over many frames for higher signal-to-noise.

**Figure 2:** (a) Schematic of an orthorhombic film grown on a (110) oriented orthorhombic substrate. The lattice parameters of the orthorhombic substrate are indicated by subscript "s". An expanded 2×2×2 pseudocubic unit cell is marked in the film, with arrows tracing the traditional *a* and *b* orthorhombic lattice parameters, denoted by subscript "f". The angle between these two directions are denoted by γ. (b) Unit cell schematic of the two cross-section views of NdNiO$_3$, along with simulated and experimental PACBED patterns of orthorhombic *Pbnm* films, showing similar features.

**Figure 3:** (a) Simulated LA-PACBED patterns for bulk NdNiO$_3$, and experimental LA-PACBED patterns for the NdGaO$_3$ substrate and the NdNiO$_3$ film on NdGaO$_3$ at room temperature. (b) LA-PACBED patterns from (a) after a Sorbel edge filter, with enhanced contrast to allow for identification of the diffraction discs. Selected diffraction discs are indexed in orthorhombic notation. The insets show the experimental 200 discs. The NdGaO$_3$ substrate shows very similar features and intensities as the simulation, while the NdNiO$_3$ film shows a different symmetry than the substrate and bulk structure. (c) Schematic of the expanded 2×2×2 pseudocubic unit cell for bulk $NdNiO_3$ and tensile strained $NdNiO_3$ films, showing relationships between key features in the orthorhombic lattice parameters and most probable Glazer octahedral tilts.

**Figure 4:** Low temperature LA-PACBED patterns from the $NdGaO_3$ substrate and $NdNiO_3$ film. While the substrate pattern is similar to its room temperature counterpart, the low temperature film displays a different symmetry from both the room temperature film and the substrate.

**Figure 5:** (a) LA-PACBED patterns of $NdNiO_3$ grown on $YAlO_3$ at room temperature and at 105 K. The patterns do not show any structural change. (b) Expanded pseudocubic unit cell schematic of the compressively strained $NdNiO_3$ film and key lattice parameters.

**Figure 1**

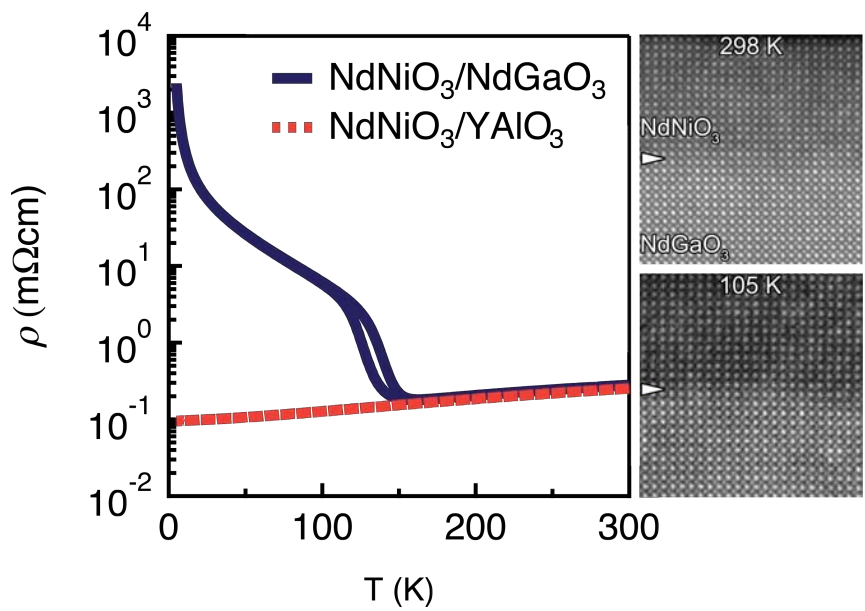

**Figure 2**

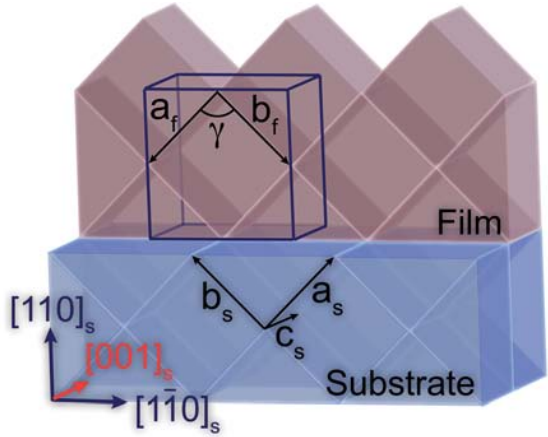

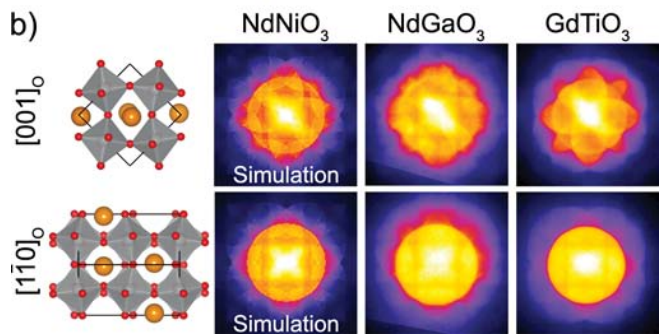

**Figure 3**

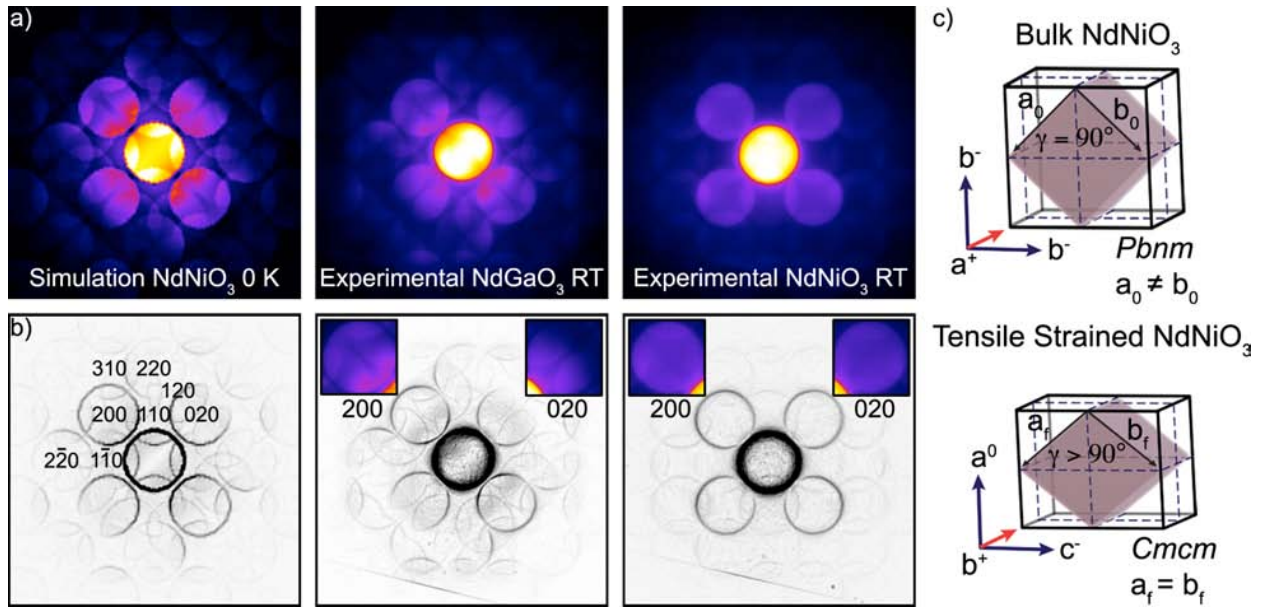

**Figure 4**

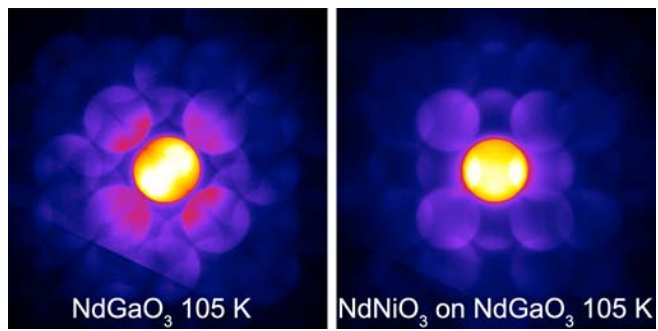

**Figure 5**

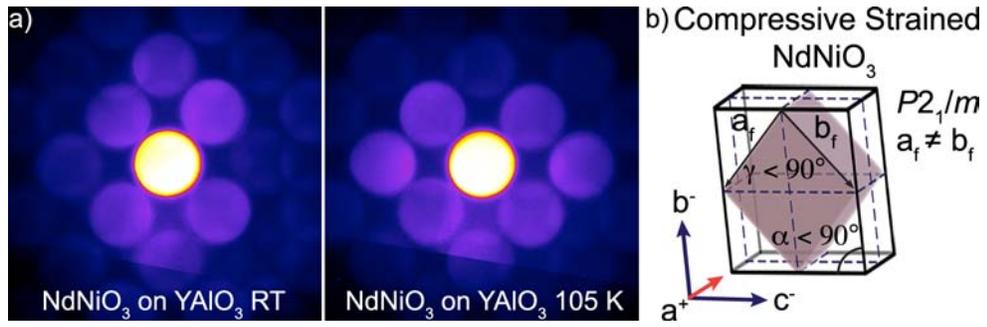